\documentclass[aps, prl, 10pt, twocolumn, superscriptaddress,floatfix]{revtex4-1}
\usepackage{bbold}
\usepackage{graphicx}
\usepackage{amsmath,amssymb,amsfonts}
\usepackage[percent]{overpic}
\usepackage{braket}
\usepackage{bm}
\usepackage[breaklinks=true,colorlinks,citecolor=blue,linkcolor=blue,urlcolor=blue]{hyperref}
\usepackage[usenames, dvipsnames]{color}
\usepackage{mathrsfs}
\begin{document}
\title{Quantum advantage in the charging process of Sachdev-Ye-Kitaev batteries}
\author{Davide Rossini}
\email{davide.rossini@unipi.it}
\affiliation{Dipartimento di Fisica dell'Universit\`a di Pisa, Largo Bruno Pontecorvo 3, I-56127 Pisa, Italy}
\affiliation{INFN, Sezione di Pisa, Largo Bruno Pontecorvo 3, I-56127 Pisa, Italy}
\author{Gian Marcello Andolina}
\email{gian.andolina@sns.it}
\affiliation{NEST, Scuola Normale Superiore, I-56126 Pisa,~Italy}
\affiliation{Istituto Italiano di Tecnologia, Graphene Labs, Via Morego 30, I-16163 Genova,~Italy}
\author{Dario Rosa}
\affiliation{School of Physics, Korea Institute for Advanced Study, 85 Hoegiro Dongdaemun-gu, Seoul 02455,~Republic of Korea}
\author{Matteo Carrega}
\affiliation{NEST, Istituto Nanoscienze-CNR and Scuola Normale Superiore, I-56127 Pisa,~Italy}
\author{Marco Polini}
\affiliation{Dipartimento di Fisica dell'Universit\`a di Pisa, Largo Bruno Pontecorvo 3, I-56127 Pisa, Italy}
\affiliation{School of Physics \& Astronomy, University of Manchester, Oxford Road, Manchester M13 9PL, United Kingdom}
\affiliation{Istituto Italiano di Tecnologia, Graphene Labs, Via Morego 30, I-16163 Genova,~Italy}
\begin{abstract}
  The exactly-solvable Sachdev-Ye-Kitaev (SYK) model has recently received considerable attention in both condensed matter and high energy physics because it describes quantum matter without quasiparticles, while being at the same time the holographic dual of a quantum black hole. In this Letter, we examine SYK-based charging protocols of quantum batteries with $N$ quantum cells. Extensive numerical calculations based on exact diagonalization for $N$ up to $16$ strongly suggest that the optimal charging power of our SYK quantum batteries displays a super-extensive scaling with $N$ that stems from genuine quantum mechanical effects. While the complexity of the non-equilibrium SYK problem involved in the charging dynamics prevents us from an analytical proof, we believe that this Letter offers the first (to the best of our knowledge) strong numerical evidence of a quantum advantage occurring due to the maximally-entangling underlying quantum dynamics.
\end{abstract}
% -------------------------------------------------------------------

\maketitle

{\it Introduction}.---In the era of quantum supremacy for quantum computing~\cite{NielsenChuang,arute_nature_2019}, research on the potential usefulness of quantum mechanical resources (such as entanglement) in energy science has led a consistent number of authors to introduce and study ``quantum batteries" (QBs). A QB~\cite{Alicki13,Campaioli18} is a system composed of $N$ identical {\it quantum cells}, where energy is stored and from which work can be extracted. 

In 2013, Alicki and Fannes~\cite{Alicki13} suggested that ``entangling unitary controls", i.e.~unitary operations acting globally on the state of the $N$ quantum cells, lead to better work extraction capabilities from a QB, when compared to unitary operations acting on each quantum cell separately. Hovhannisyan et al.~\cite{Hovhannisyan13} were the first to demonstrate that entanglement generation leads to a speed-up in the process of work extraction, thereby leading to larger delivered power. Later on, the authors of Refs.~\cite{Binder15, Campaioli17} focussed on the charging (rather than the discharging) procedure and identified two types of charging schemes: i)
the {\it parallel} charging scheme in which each of the $N$ quantum cells is acted upon independently of the others; and ii) the {\it collective} charging scheme, where global unitary operations (i.e.~the entangling unitary controls of Ref.~\cite{Alicki13}) acting on the full Hilbert space of the $N$ quantum cells are allowed. They were able to show that, in the collective charging case and for $N\geq 2$, the charging power of a QB is larger than in the parallel scheme.
This collective speed-up (stemming from entangling operations) during the charging procedure of a QB has been named ``quantum advantage".
\begin{figure}[t]
\centering
\vspace{1.em}
\begin{overpic}[width=1 \columnwidth]{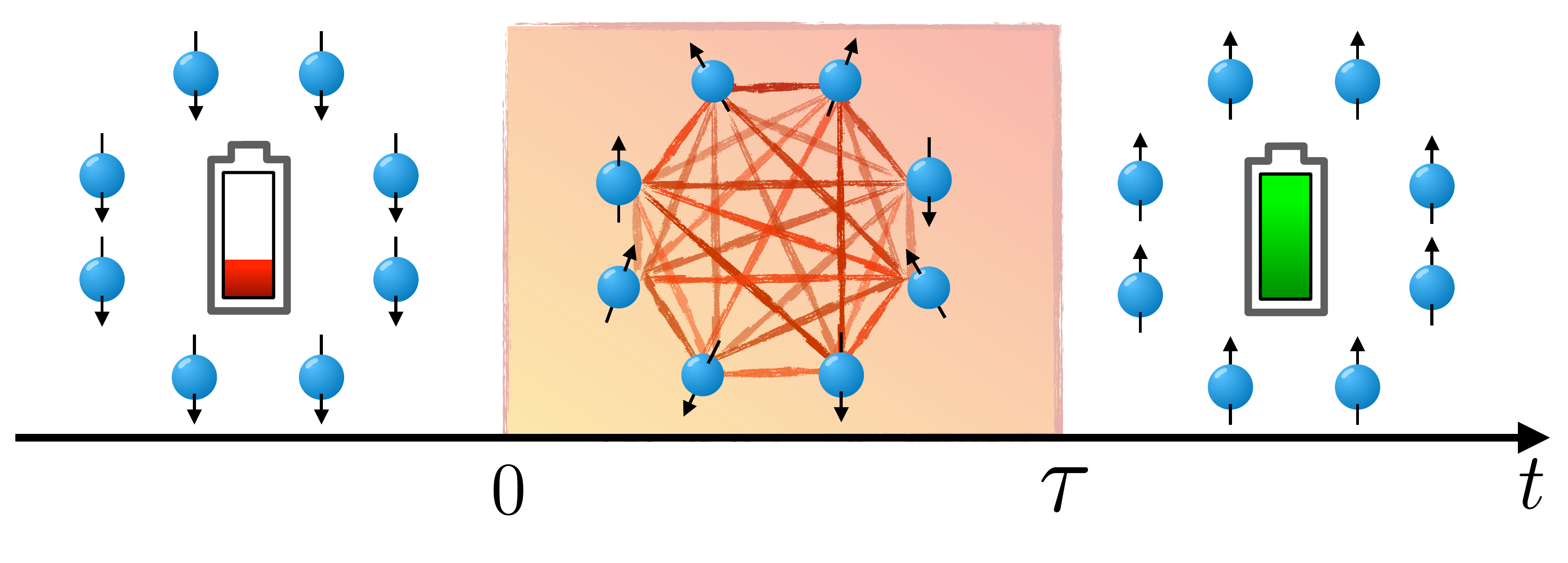}\put(6,73){ }\end{overpic}
\caption{The charging protocol of a QB made of $N$ spin-$1/2$ units, described by the $\mathcal{\hat H}_0$ in Eq.~(\ref{eq:H_NonInt}). At time $t < 0$, the battery is fully discharged. In the time interval $0 < t < \tau$, the interacting charging Hamiltonian $\mathcal{\hat H}_1$ is switched on, and energy is injected via the quench. Finally, at time $\tau$, interactions are switched off and $\mathcal{\hat H}_0$ is switched back on, so that the stored energy $E_{N}(\tau)$ is conserved thereafter.  \label{fig:Protocol}}
\end{figure}

In the quest for such quantum advantage and potential laboratory implementations of QBs---based, e.g., on circuit quantum electrodynamics and trapped-ion setups---the abstract concepts of ``quantum cell" and ``entangling operations" have been recently spelled out more explicitly~\cite{Le17,Ferraro17,Andolina18,Andolina19,Farina18,Zhang18,zhang_arxiv_2018,Barra19,Rossini19,Ghosh19,Santos19,Chen19,Monsel19,Caravelli19,Pintos19,Gherardini19,Kamin19,Pirmoradian19,Farre18,Andolina18c}. Different prototypes of QBs have been devised: i) Dicke models, where arrays of $N$ qubits (i.e.~the proper battery) are coupled to a harmonic energy source~\cite{Ferraro17,Andolina18,Andolina19,Farina18,Zhang18,zhang_arxiv_2018}; ii) deterministic spin chains~\cite{Le17,Farre18,Andolina18c}; and iii) disordered spin chains~\cite{Rossini19,Ghosh19}. These quantum cells can be charged by switching on either direct~\cite{Le17,Rossini19,Ghosh19} or effective~\cite{Ferraro17,Andolina18,Andolina19,Farina18,Zhang18,zhang_arxiv_2018} interactions between them. 

The authors of Refs.~\cite{Ferraro17, Le17} proposed two concrete implementations of the collective charging scheme, and claimed the existence of a quantum advantage over the parallel charging procedure. However, Juli\`a-Farr\'e et al.~\cite{Farre18} noticed that the Hamiltonians adopted in Refs.~\cite{Le17, Ferraro17} were not properly defined in the thermodynamic limit, in the sense that their average values did not display extensivity with $N$, but, rather, displayed a super-linear growth with $N$. Moreover, the same authors were able to derive a rigorous bound for the charging power, allowing to distinguish between a genuine entanglement-induced speed-up and spurious effects, given e.g. by the lack of a well-defined thermodynamic limit. In agreement with Ref.~\cite{Andolina18c}, the conclusion of Ref.~\cite{Farre18} is that {\it all} the many-body QB models proposed in the literature so far do not feature any genuine quantum advantage.

Motivated by this literature, we propose a model of a QB which i) is properly defined in the thermodynamic limit and ii) unequivocally presents a genuine quantum advantage. Our implementation relies on the Sachdev-Ye-Kitaev (SYK) model~\cite{Sachdev93,Kitaev15,gu_2019,rosenhaus_jpa_2019}, which has recently attracted a great deal of attention for its exact solvability and profound properties.
The SYK model describes quantum matter with no quasiparticles. It displays fast scrambling~\cite{Maldacena16, Roberts18}, has a nonzero entropy density at vanishing temperature~\cite{Sachdev15,Georges01}, all its eigenstates exhibit volume-law entanglement entropy~\cite{Liu18,Huang19}, and is holographically connected
to the dynamics of ${\rm AdS}_2$ horizons of quantum black holes~\cite{Kitaev15,gu_2019,Sachdev10,Sachdev10b}. Proposals to realize the SYK Hamiltonian have been recently put forward and rely on ultra-cold atoms~\cite{Danshita17},
graphene flakes with irregular boundaries~\cite{Chen18}, and topological superconductors~\cite{Chew17,Pikulin17}.  

{\it Many-body QBs and figures of merit.}---Consider a QB made of $N$ identical quantum cells (for a cartoon, see Fig.~\ref{fig:Protocol}), which are governed by the following free and local Hamiltonian ($\hbar = 1$):
\begin{equation}\label{eq:H_NonInt}
\mathcal{\hat H}_0 = \sum_{j=1}^N \hat h_j~.
\end{equation}
At time $t=0$, the system is prepared in its ground state $|0\rangle$, physically representing the discharged battery.
By suddenly switching on a suitable interaction Hamiltonian $\mathcal{\hat H}_1$ for a finite amount of time $\tau$ (and switching off $\mathcal{\hat H}_0$),
one aims at injecting as much energy as possible into the quantum cells~\cite{Binder15,Campaioli17,Le17}. The time interval $\tau$ is called the {\it charging time} of the protocol. The full model Hamiltonian can be thus written as
\begin{equation}
\label{eq:protocol}
\mathcal{\hat H}(t) = \mathcal{\hat H}_{\rm 0} + \lambda(t) \big(\hat{\cal H}_1 - \hat{\cal H}_{\rm 0} \big)~,
\end{equation}
where $\lambda(t)$ is a classical parameter that represents the external
control exerted on the system, and which is assumed to be given by
a step function equal to $1$ for $t\in[0,\tau]$ and zero elsewhere. Such charging protocol is experimentally feasible, e.g. in cold-atom setups \cite{Polkovnikov11}, where implementing sudden quenches is a standard procedure.
Accordingly, denoting by $|\psi(t) \rangle$ the state of the system at time $t$, its total energy
$E^{\rm tot}_N(t) = {\langle} \psi(t) |\mathcal{\hat H}(t)| \psi(t) \rangle$
is constant for all values of $t$ but $t=0$ and $t=\tau$ (the switching points).

The energy injected into the $N$ quantum cells can be expressed in terms of the mean local energy
at the end of the protocol, $E_N(\tau) =
\braket{\psi(\tau) |\mathcal{\hat H}_0| \psi(\tau)}$. In writing the previous equation, we have set to zero the ground-state energy $\braket{0|\hat{\cal H}_0|0}$.
Other crucial figures of merit are the average charging power $P_N(\tau)=E_N(\tau) / \tau$ and its optimal value
\begin{equation}\label{eq:Power}
P_N({\tau}^*)=\max_{\tau>0} P_N({\tau})~,
\end{equation}
obtained at time $\tau^*$. In the following, we will be mainly interested in the scaling
of the optimal charging power $P_N({\tau}^*)$ with the number $N$ of quantum cells.

{\it SYK-based charging protocols.}---We assume each quantum cell to be a spin-$1/2$ system. In the absence of charging operations, the system is described by the
non-interacting Hamiltonian~\eqref{eq:H_NonInt}, with  $\hat h_j =\omega_0 \hat \sigma^y_j / 2$.
Here, $\omega_0>0$ represents a magnetic field strength (with units of energy)
and $\hat \sigma^\alpha_j$ ($\alpha=x,y,z$) are the Pauli matrices. The battery energy $E_N(\tau)$ will be measured in units of the energy scale $\omega_{0}$. 
At time $t=0$, the quantum cells are initialized in the ground state of $\mathcal{\hat H}_0$, $\ket{0} = \bigotimes_{j=1}^{N} \ket{\downarrow^{(y)}}_{j}$, where $\hat \sigma^y_j \ket{\downarrow^{(y)}}_{j} = -\ket{\downarrow^{(y)}}_{j}$. 

For the charging Hamiltonian $\mathcal{\hat H}_1$, we use the complex SYK (c-SYK)~\cite{gu_2019,Fu16,Davison17} model Hamiltonian:
\begin{equation}\label{eq:HamB}
\mathcal{\hat H}^{\textrm{c-SYK}}_{1} = \sum_{i,j,k,l=1}^N J_{i,j,k,l}\hat{c}^\dagger_i \hat{c}^\dagger_j \hat{c}_k \hat{c}_l~,
\end{equation}
where $\hat{c}^\dagger_j$ ($\hat{c}_j$) is a spinless fermionic creation (annihilation) operator~\cite{majorana}. This has to be understood in its
spin-$1/2$ representation, which is obtained by the Jordan-Wigner (JW) transformation $\hat{c}^\dagger_j= \hat \sigma^{+}_j \big( \Pi_{m=1}^{j-1} \hat \sigma_m^{z} \big)$, where $\hat{\sigma}^{\pm}_j\equiv( \hat{\sigma}^{x}_j \pm i \hat{\sigma}^{y}_j)/2$~\cite{SM}.
The couplings $J_{i,j,k,l}$ are zero-mean Gaussian-distributed complex random variables,
with variance $\langle \! \langle J^2_{i,j,k,l} \rangle \! \rangle = J^2/N^3$,
satisfying $J_{i,j,k,l} =J^*_{k,l,i,j}$ and $J_{i,j,k,l} =-J_{j,i,k,l}=-J_{i,j,l,k}$. 
In the following, we average any quantity of interest $\mathcal{O}$ over the distribution of $\{ J_{i,j,k,l}\}$,
and denote by $\langle \! \langle \mathcal{O} \rangle \! \rangle$ the averaged value, i.e.~$\langle \! \langle \mathcal{O} \rangle \! \rangle\equiv \int \! P(\{J_{i,j,k,l} \}) \mathcal{O}(\{J_{i,j,k,l}\}) \; {\rm d}\{ J_{i,j,k,l}\}$.
\begin{figure*}[t]
  \centering
  \vspace{1.em}
  \vspace{1.em}
  \begin{overpic}[width=0.68\columnwidth]{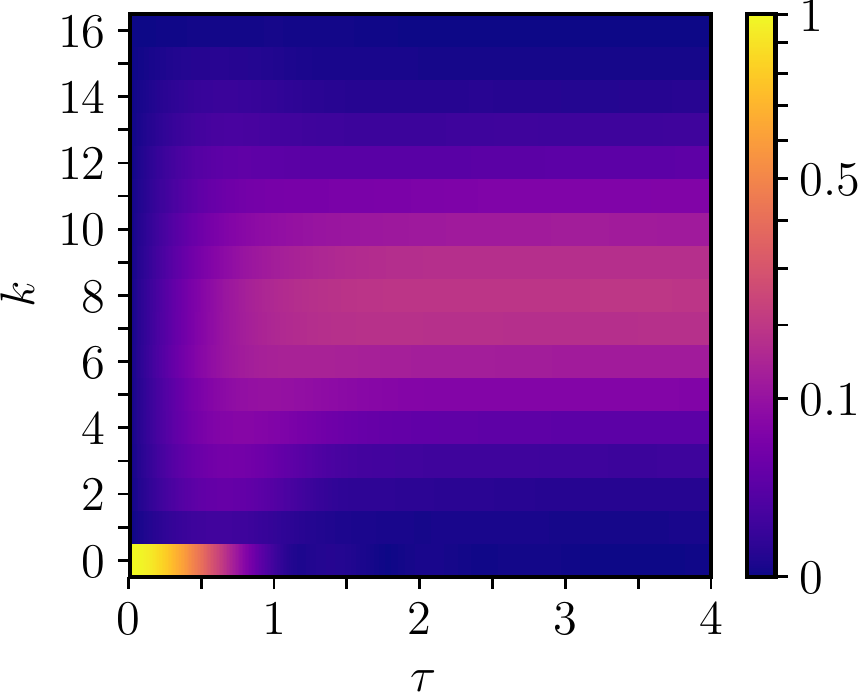}\put(0,80){\normalsize (a)}\end{overpic}
  \begin{overpic}[width=0.68\columnwidth]{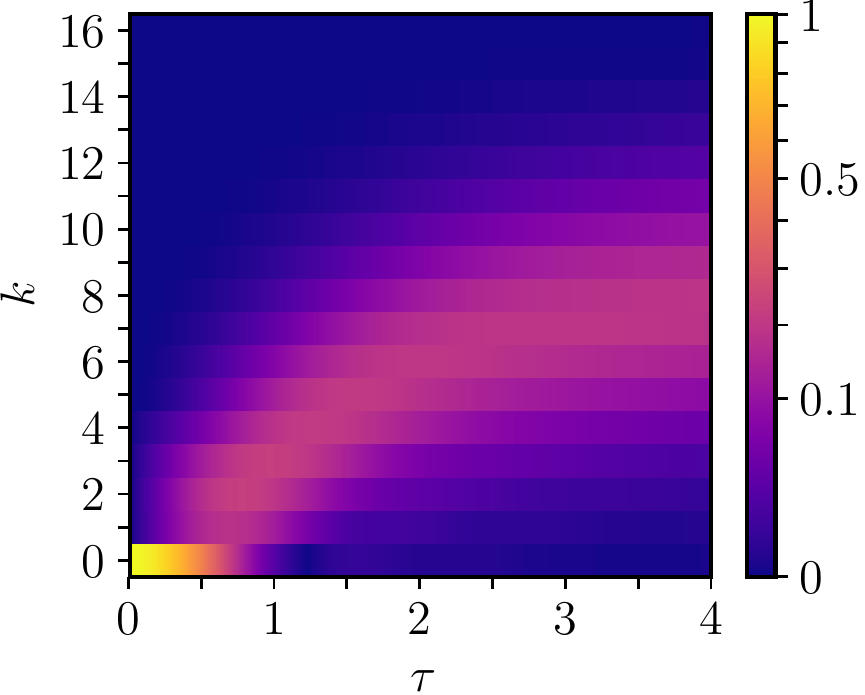}\put(0,80){\normalsize (b)}\end{overpic}
  \begin{overpic}[width=0.68\columnwidth]{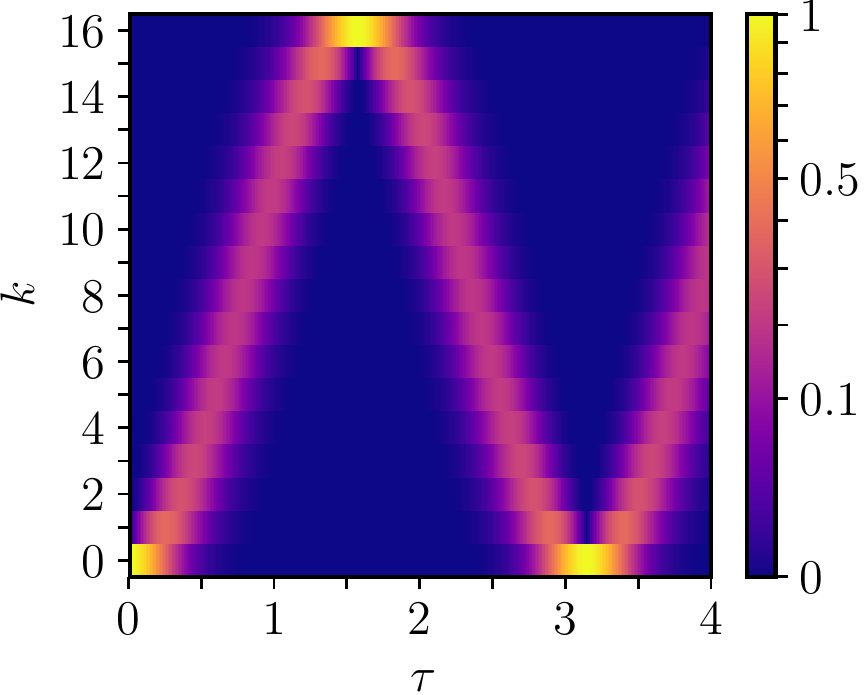}\put(0,80){\normalsize (c)}\end{overpic}
  \caption{Dynamics of the dimensionless population $p_k(\tau)$ of the QB energy levels as a function of time $\tau$ (in units of $1/J$) and the level index $k$ for three different charging protocols: c-SYK (a), b-SYK with $\bar{J}=J$ (b), and parallel with $K=J$  (c). Data in panels (a) and (b) correspond to a single realization of disorder in the couplings $J_{i,j,k,l}$ and $\bar{J}_{i,j,k,l}$.\label{fig:Levels}}
\end{figure*}

We emphasize that our choice of battery and charging Hamiltonians is such that $[\mathcal{\hat H}_0, \mathcal{\hat H}_1] \neq 0$,
a condition which ensures energy injection into the QB by the charging protocol~\eqref{eq:protocol}. Note, finally, that the Hamiltonian in Eq.~\eqref{eq:HamB} is invariant under particle-hole symmetry (PHS) in the thermodynamic limit $N\to \infty$. Extra terms, however, need to be added to it in order to enforce PHS at any finite $N$~\cite{Fu16}:
\begin{align}\label{eq:HamBPHS}
\mathcal{\hat H}^{\textrm{c-SYK (PHS)}}_{1} & = \mathcal{\hat H}^{\textrm{c-SYK}}_{1} + \frac{1}{2} \sum_{i,j,k,l=1}^N J_{i,j,k,l} \\
  & \times \big( \delta_{i,k} \hat{c}^\dagger_j \hat{c}_l - \delta_{i,l}\hat{c}^\dagger_j \hat{c}_k - \delta_{j,k} \hat{c}^\dagger_i \hat{c}_l
  + \delta_{j,l} \hat{c}^\dagger_i \hat{c}_k \big)~. \nonumber 
\end{align}
Hereafter, we will always use this version of the c-SYK model.
We have however checked that our main findings do not qualitatively change if PHS is not enforced and (\ref{eq:HamB}), rather than (\ref{eq:HamBPHS}), is used as charging Hamiltonian.

In the following, we will also consider charging Hamiltonians based on a bosonic version of the SYK model (b-SYK)~\cite{Fu16}:
\begin{equation} \label{eq:HamBose}
  \mathcal{\hat H}^{\textrm{b-SYK}}_{1} = \sum_{i,j,k,l=1}^N \bar{J}_{i,j,k,l} \hat{b}^\dagger_i \hat{b}^\dagger_j \hat{b}_k \hat{b}_l~,
\end{equation}
where $\hat{b}^\dagger_j$ ($\hat{b}_j$) creates (annihilates) an hard-core boson. The following relations are obeyed: $\{\hat{b}_j,\hat{b}^\dagger_j\}=1$
and $[\hat{b}_i,\hat{b}_j]=0$ for $i\neq j$.
Hence, $\hat{b}^\dagger_j$ can be directly written in its spin representation as $\hat{b}^\dagger_j= \sigma_j^{+}$.  Similarly to $J_{i,j,k,l}$, the quantities $\bar{J}_{i,j,k,l}$ in Eq.~(\ref{eq:HamBose}) are random, Gaussian-distributed variables, with variance $\langle \! \langle \bar{J}^2_{i,j,k,l} \rangle \! \rangle = J^2/N^3$, satisfying $\bar{J}_{i,j,k,l} = \bar{J}^*_{k,l,i,j}$ and $\bar{J}_{i,j,k,l} = \bar{J}_{j,i,k,l} = \bar{J}_{i,j,l,k}$ (in order to comply with the bosonic commutation rules of the model). For PHS to hold, we enforce the site indices $i,j,k,l$ in Eq.~(\ref{eq:HamBose}) to be all different~\cite{Fu16}.
 Note that the dependence of the variance of the couplings $J_{i,j,k,l}$ and $\bar{J}_{i,j,k,l}$ on the inverse third power of $N$ ensures that all our SYK charging Hamiltonians are well-defined in the thermodynamic limit. Indeed, their average values scale extensively with $N$~\cite{GarciaGarcia}.

Finally, we will also examine a parallel charging protocol~\cite{Binder15,Campaioli17} based on the following Hamiltonian
\begin{equation}\label{eq:HamParallel}
\mathcal{\hat H}^{\|}_{1} = K \sum_{j=1}^N \hat \sigma^x_j~.
\end{equation}
In this case, each of the $N$ quantum cells is acted upon independently of the others and no entanglement is generated~\cite{Farre18}. The charging protocol based on $\mathcal{\hat H}^{\|}_{1}$  will therefore serve as reference model, to be compared against c-SYK and b-SYK charging models.

{\it Microscopy of the charging dynamics in energy space.}---As an indicator of the speed of the dynamics, we start by looking at the time evolution of the energy-level occupations. Consider the spectral decomposition of Hamiltonian~\eqref{eq:H_NonInt}:  $\mathcal{\hat H}_0 = \sum_{k=0}^N \epsilon_k \sum_i \ket{k,i}\bra{k,i}$, where $\epsilon_k = k\omega_0$ denote its eigenvalues and
the index $i$ accounts for the degenerate eigenvectors. We are interested in the dynamics of the populations:
\begin{equation}
p_k(\tau)=\sum_i | \braket{{k,i} |\psi(\tau)}|^2~.
\end{equation}
Figure~\ref{fig:Levels} displays $p_k(\tau)$ for the three charging Hamiltonians mentioned above: c-SYK (a), b-SYK (b), and parallel (c).
While in the latter two cases the charging protocol generates a dynamics that is clearly local in energy space,
this is not the case for the c-SYK model. This charging model generates a non-local population dynamics in energy space, which manifests as a sudden macroscopic population of excited levels. Indeed, after an ultrashort ``thermalization" time~\cite{Eberlein17} , a central band of excited energy levels appears uniformly populated. (Further details on the thermalization properties of c-SYK QBs are provided in Ref.~\cite{SM}).
 This non-locality is a direct realization of the global charging dynamics envisioned by the authors of Ref.~\cite{Binder15}. Recurrences appearing in the charging dynamics highlighted in panel (c) witness the integrability of the parallel Hamiltonian in Eq.~\eqref{eq:HamParallel}, which is absent in the SYK models.

{\it Power, bounds, and quantum advantage.}---Quantitative conclusions on the charging performances of SYK QBs, compared to those of other 
reference many-body QBs, can be drawn from the analysis of the optimal power $P_N(\tau^*)$ in Eq.~\eqref{eq:Power} and its scaling with $N$. 
Specifically, a rigorous certification of the quantum origin of the charging advantage of the c-SYK model
can be achieved by considering the following bound~\cite{Farre18}:
\begin{equation}\label{eq:Power Bound}
P_N(\tau) \leq 2 \sqrt {\Delta_\tau \mathcal{\hat H}^2_0 ~\Delta_\tau \mathcal{\hat H}^2_1}~,
\end{equation}
where $\Delta_\tau \mathcal{\hat H}^2 \equiv (1/\tau) \int_0^\tau dt [ \braket{\mathcal{\hat H}^2}_t - (\braket{ \mathcal{\hat H}}_t)^2]$
and $\braket{ \mathcal{\hat O} }_t\equiv \braket{{\psi(t)}| \mathcal{\hat O} |{\psi(t)}}$. Here, $\Delta_\tau \mathcal{\hat H}^2_1$ represents the charging speed in the Hilbert space:
larger values of such quantity correspond to trivial increases of the charging speed.
In contrast, $\Delta_\tau \mathcal{\hat H}^2_0$ is connected with the distance traveled in the Hilbert space.
An enhancement of it can be linked to shortcuts in the Hilbert space: starting from a pure state and
going through highly entangled states, it is possible to reduce the length of the trajectory in such space,
consequently enhancing the charging power~\cite{Farre18}.
This is a genuine quantum effect, with no classical analogue.
Any increase of the average optimal power linked to $\Delta_\tau \mathcal{\hat H}^2_0$ 
can be considered as the smoking gun of a genuine quantum advantage, unreproducible by classical dynamics.
A detailed derivation of the bound~\eqref{eq:Power Bound} is provided in Ref.~\cite{SM}.

\begin{figure}[t]
  \vspace{1.em}
  \begin{overpic}[width=0.85\columnwidth]{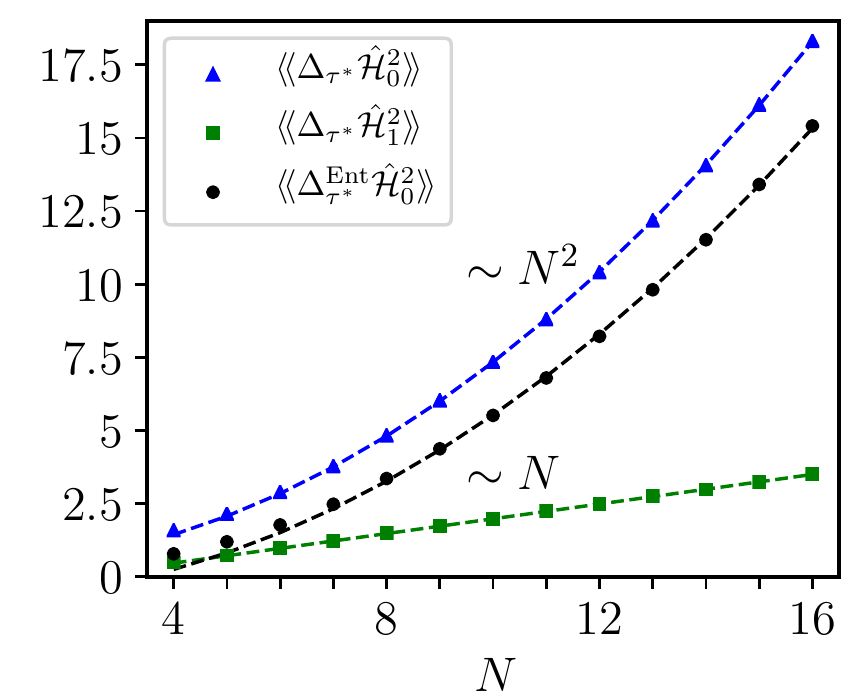}\put(0,80){\normalsize (a)}\end{overpic} \\
  \vspace*{0.2cm}
  \begin{overpic}[width=0.85\columnwidth]{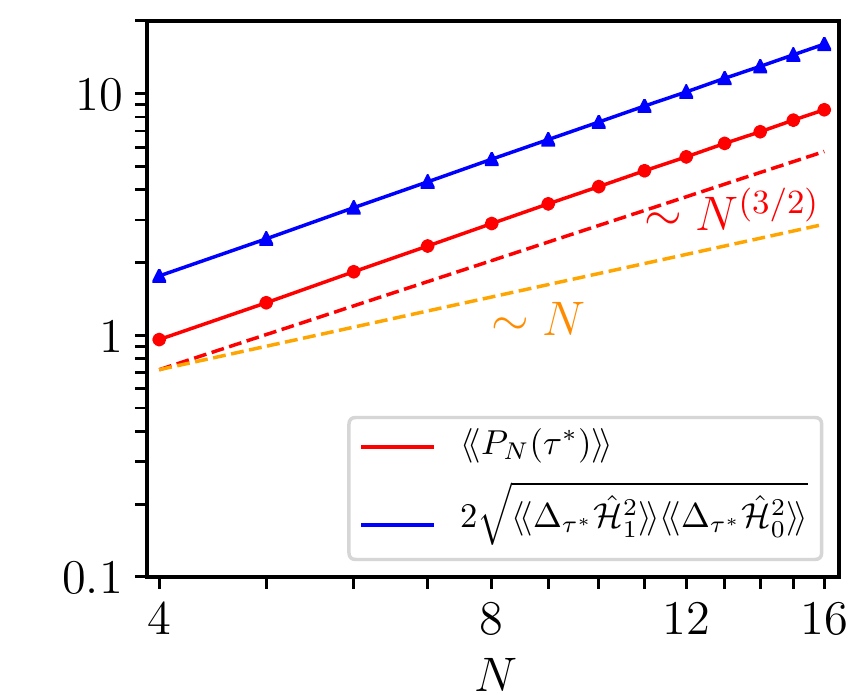}\put(0,80){\normalsize (b)}\end{overpic} 
  \caption{Panel (a) The relevant quantities for the bound~\eqref{eq:Power Bound2}, evaluated at the optimal time $\tau^*$, and averaged over disorder: time-averaged variances $\langle \! \langle \Delta_{\tau^*} \mathcal{\hat H}^2_0 \rangle \! \rangle$ (blue triangles, in units of $\omega^2_{0}$), $\langle \! \langle \Delta_{\tau^*} \mathcal{\hat H}^2_1 \rangle \! \rangle$ (green squares, in units of $J^2$), $\langle \! \langle \Delta^{\rm Ent}_{\tau^*} \mathcal{\hat H}^2_0 \rangle \! \rangle$ (black circles, in units of $\omega^2_{0}$), as functions of $N$. Dashed curves denote linear (green) and quadratic (blue, black) fits to the numerical results. The four data points corresponding to the smallest $N$ have been discarded from the fits. Panel (b) The optimal power (red) $\langle\!\langle P_{N}(\tau^*) \rangle\!\rangle$ and the quantity in the right-hand-side of Eq.~\eqref{eq:Power Bound2} (blue) are plotted as functions of $N$, in a log-log scale and in units of $\omega_{0}J$. Dashed lines correspond to power laws $\sim N^{1+k}$ ($k=0.5$: red; $k=0$: orange) and are plotted as guides to the eye. Data in this figure refer to the c-SYK QB model, and have been obtained after averaging over $N_{\rm dis} = 10^3$ (for $N=4, \ldots, 10$), $5 \times 10^2$ (for $N=11, 12$), and $10^2$ (for $N=13, \ldots, 16$) instances of disorder in the couplings $\{ J_{i,j,k,l} \}$.}
  \label{fig:Power}
\end{figure}

If the battery Hamiltonian $\mathcal{\hat H}_0$ is made of a sum of local terms, as in Eq.~\eqref{eq:H_NonInt},
it is possible to write $\Delta_\tau \mathcal{\hat H}^2_0$ as: $\Delta_\tau \mathcal{\hat H}^2_0 = \Delta^{\rm Loc}_\tau \mathcal{\hat H}^2_0 + \Delta^{\rm Ent}_\tau \mathcal{\hat H}^2_0$,
with~\cite{Farre18}
\begin{eqnarray}
  \label{eq:BoundEnt}
  \Delta^{\rm Loc}_\tau \mathcal{\hat H}^2_0 & \equiv &
  \frac{1}{\tau} \int_0^\tau dt \sum_i \Big[ \braket{\hat h_i^2}_t - \braket{\hat h_i}^2_t \Big]~, \label{eq:DeltaH0} \\
  \Delta^{\rm Ent}_\tau \mathcal{\hat H}^2_0~ & \equiv & \frac{1}{\tau} \int_0^\tau dt
  \sum_{i\neq j} \Big[ \braket{\hat h_i \hat h_j}_t - \braket{\hat h_i}_t \braket{\hat h_j}_t \Big]~. \label{eq:DeltaEntH0}
\end{eqnarray}
The quantity~\eqref{eq:DeltaH0}, being a sum of local terms, scales linearly with $N$ (i.e.~is extensive) by construction.
On the other hand, $\Delta^{\rm Ent}_\tau \mathcal{\hat H}^2_0$, whose explicit form can be immediately linked to correlations between sites $i$ and $j$, may display a super-linear scaling with $N$. Due to the non-linearity of the bound~\eqref{eq:Power Bound}, which applies to a single disorder realization, averaging over disorder is not straightforward.
Through the Cauchy-Schwarz inequality, though, it is possible to rewrite it as
$\langle \! \langle P_N(\tau) \rangle \! \rangle \leq
2 \, \Big\langle \!\! \Big\langle \sqrt{\Delta_\tau \mathcal{\hat H}^2_0 \, \Delta_\tau \mathcal{\hat H}^2_1} \Big\rangle \!\! \Big\rangle \leq
2 \sqrt { \langle \! \langle \Delta_\tau \mathcal{\hat H}^2_0 \rangle \! \rangle \langle \! \langle \Delta_\tau \mathcal{\hat H}^2_1 \rangle \! \rangle}$,
meaning that one can separately study the averaged quantities
$\langle \! \langle \Delta_\tau \mathcal{\hat H}^2_0 \rangle \! \rangle$ and
$\langle \! \langle \Delta_\tau \mathcal{\hat H}^2_1 \rangle \! \rangle$.
Here we are interested in the scaling at the optimal time $\tau^*$, thus we focus on
\begin{equation}\label{eq:Power Bound2}
\langle \! \langle P_N(\tau^*) \rangle \! \rangle  \leq   2 \sqrt { \langle \! \langle \Delta_{\tau^*} \mathcal{\hat H}^2_0 \rangle \! \rangle\langle \! \langle \Delta_{\tau^*} \mathcal{\hat H}^2_1 \rangle \! \rangle}~.
\end{equation}
Since the battery energy is measured in units of $\omega_{0}$ and time in units of $1/J$, the averaged charging power $\langle\!\langle P_{N}(\tau^*) \rangle\!\rangle$ is measured in units of $\omega_{0} J$. Given this choice, we specify the energy scales of the b-SYK and parallel-charging protocols by setting $\bar{J} = K = J$~\cite{rescaling}.

Figure~\ref{fig:Power}(a) shows the relevant quantities for the bound~\eqref{eq:Power Bound2}, for a c-SYK QB.
While $\langle \! \langle \Delta_{\tau^*} \mathcal{\hat H}^2_1 \rangle \! \rangle$ is extensive in $N$,
we observe that both $\langle \! \langle \Delta_{\tau^*} \mathcal{\hat H}^2_0 \rangle \! \rangle$
and $\langle \! \langle \Delta^{\rm Ent}_{\tau^*} \mathcal{\hat H}^2_0 \rangle \! \rangle$ display a super-linear scaling with $N$, which is compatible with a quadratic growth. This means that, during the time evolution, the c-SYK charging Hamiltonian generates the maximum possible non-locality between the quantum cells, in the form of $N$-partite entanglement~\cite{Farre18}. This, together with Eq.~\eqref{eq:Power Bound2}, suggests a super-linear scaling with $N$ of the optimal charging power,
\begin{equation}
  \langle\!\langle P_{N}(\tau^*) \rangle\!\rangle \sim N^{1+k}~, \quad~{\rm with}~k > 0~,
\end{equation}
where $k \approx 0.5$. For the first time in the literature on QB models~\cite{Le17,Ferraro17,Andolina18,Andolina19,Farina18,Zhang18,zhang_arxiv_2018,Barra19,Rossini19,Ghosh19,Santos19,Chen19,Monsel19,Caravelli19,Pintos19,Gherardini19,Kamin19,Farre18,Andolina18c,Pirmoradian19}, we are thus in a situation where the power enhancement is linked to $\Delta_\tau \mathcal{\hat H}^2_0$, a fact that hints at a quantum advantage (i.e.~advantage over any classical battery) displayed by the c-SYK model with respect to the charging task. Further details on the comparison between quantum and classical many-body batteries are given in Ref.~\cite{SM}.

The left- and right-hand-side members of the inequality~\eqref{eq:Power Bound2} are displayed in Fig.~\ref{fig:Power}(b), red and blue data, respectively. We clearly see a super-linear scaling with $N$ ($k = 0.5$ corresponds to the red dashed straight line). We have also considered the b-SYK and parallel-charging models, showing that, in both cases, all the quantities $\langle \! \langle \Delta_{\tau^*} \mathcal{\hat H}^2_0 \rangle \! \rangle$,
$\langle \! \langle \Delta_{\tau^*} \mathcal{\hat H}^2_1 \rangle \! \rangle$, and
$\langle \! \langle \Delta^{\rm Ent}_{\tau^*}  \mathcal{\hat H}^2_0 \rangle \! \rangle$ are extensive in $N$~\cite{SM}.
In agreement with the results shown in Figs.~\ref{fig:Levels}-\ref{fig:Power}, we thus conclude that these two QB models do not display any genuine quantum advantage.

We finally recall that optimal charging powers that scale faster than $N$ have been found in Refs.~\cite{Le17,Ferraro17}. Unfortunately, such super-linear scalings do not stem from $\Delta_\tau \mathcal{\hat H}^2_0$ but rather from $\Delta_\tau \mathcal{\hat H}^2_1$, and therefore have no quantum origin~\cite{Farre18}. The fact that the Hamiltonians used in Refs.~\cite{Le17,Ferraro17} are not properly defined in the thermodynamic limit is ultimately at the origin of the spurious super-extensive scaling of the optimal charging power. This is {\it explicitly} shown in Ref.~\cite{SM} for Dicke QBs.
In this Letter, we have bypassed this problem by choosing the appropriate scaling~\cite{Sachdev93,Kitaev15,gu_2019,rosenhaus_jpa_2019,Fu16,Davison17} with $N$ of the variance $\langle \! \langle J^2_{i,j,k,l} \rangle \! \rangle = J^2/N^3$ of the c-SYK coupling parameters.

Alternatively, another strategy to rule out any spurious effect on the optimal charging power is to use a ``renormalization" approach that consists in dividing the charging Hamiltonian by its operator norm~\cite{SM}. This procedure allows for a fair comparison between different QB models~\cite{Campaioli17}. In agreement with the results illustrated above, we have found a clear increase with $N$ of the optimal charging power only for the renormalized c-SYK Hamiltonian~\cite{SM}.

In the future, it will be interesting to study SYK-type models in the context of heat engines~\cite{Campisi16,Holubec18}, where minimizing time scales is also of central importance.

{\it Acknowledgements.}---It is a great pleasure to thank M. Fanizza, S. Juli\`a-Farr\'e, M. Lewenstein, T. Salamon, and S. Tirone for useful conversations. D. Rosa wishes to thank  the Simons Center for Geometry and Physics (Stony Brook University, USA) and the organizers of the program ``Universality and ergodicity in quantum many-body systems'' for their support and warm hospitality. M.C. acknowledges support from the Quant-EraNet project ``Supertop".

\clearpage 
\setcounter{section}{0}
\setcounter{equation}{0}%
\setcounter{figure}{0}%
\setcounter{table}{0}%

\setcounter{page}{1}

\renewcommand{\thetable}{S\arabic{table}}
\renewcommand{\theequation}{S\arabic{equation}}
\renewcommand{\thefigure}{S\arabic{figure}}
\renewcommand{\bibnumfmt}[1]{[S#1]}
\renewcommand{\citenumfont}[1]{S#1}

\onecolumngrid

\begin{center}
\textbf{\Large Supplemental Material for:\\ ``Quantum advantage in the charging process of Sachdev-Ye-Kitaev batteries''}
\bigskip

Davide Rossini, $^{1,\,2}$
Gian Marcello Andolina,$^{3,\,4}$
Dario Rosa,$^{5}$
Matteo Carrega,$^{6}$ and
Marco Polini$^{1,\,7,\,4}$

\bigskip
$^1$\!{\it Dipartimento di Fisica dell'Universit\`a di Pisa, Largo Bruno Pontecorvo 3, I-56127 Pisa, Italy}

$^2$\!{\it INFN, Sezione di Pisa, Largo Bruno Pontecorvo 3, I-56127 Pisa, Italy}

$^3$\!{\it NEST, Scuola Normale Superiore, I-56126 Pisa,~Italy}

$^4$\!{\it Istituto Italiano di Tecnologia, Graphene Labs, Via Morego 30, I-16163 Genova,~Italy}

$^5$\!{\it School of Physics, Korea Institute for Advanced Study, 85 Hoegiro Dongdaemun-gu, Seoul 02455,~Republic of Korea}

$^6$\!{\it NEST, Istituto Nanoscienze-CNR and Scuola Normale Superiore, I-56127 Pisa,~Italy}

$^7$\!{\it School of Physics \& Astronomy, University of Manchester, Oxford Road, Manchester M13 9PL, United Kingdom}

\bigskip

In this Supplemental Material we provide additional information on the way in which the c-SYK model is mapped
onto a spin-$1/2$ model and include a few details on the procedure adopted for the numerical calculations.
We formally derive the bound in Eq.~(9) of the main text, discuss the implications of such bound for quantum and classical batteries, show explicit numerical evidence for the lack of a quantum advantage in b-SYK and parallel quantum-battery charging models, 
and discuss the behavior of the optimal power after rescaling the charging Hamiltonians with their operator norm.
We also show that the SYK model induces a fast thermalization to an infinite temperature state. Finally, we discuss the lack of a super-linear charging power in a Dicke battery, provided that thermodynamic consistency is enforced.

\end{center}

%\maketitle

%
\appendix
\twocolumngrid

\section{On the JW transformation and other details on the numerical calculations}
\label{Appendix:J-W Transformations}

The c-SYK model~\cite{Fu16S} for finite $N$ is best handled numerically after mapping it onto a spin model. This is accomplished through the JW transformation. For the sake of clarity,  we here report the c-SYK model Hamiltonian [cf.~Eq.~(4) in the main text]:
\begin{equation}\label{eq:HamBSupp}
\mathcal{\hat H}^{\textrm{c-SYK}}_{1} = \sum_{i,j,k,l=1}^N J_{i,j,k,l} \, \hat{c}^\dagger_i \, \hat{c}^\dagger_j \, \hat{c}_k \, \hat{c}_l~.
\end{equation}
Here, $\hat{c}^\dagger_j$ ($\hat c_j)$ creates (annihilates) a complex spinless fermion and the usual fermionic anticommutation relations, $\{ c^\dagger_i, c_j \} = \delta_{i,j}$, $\{ c_i, c_j \} = 0$, hold true. The JW transformation, which maps spinless fermions into spin-$1/2$ degrees of freedom, reads as following:
\begin{equation}
  \hat{c}^\dagger_j= \hat \sigma_j^{+} \Bigg[ \prod_{m=1}^{j-1} \hat \sigma_m^{z} \Bigg]~, \qquad
  \hat{c}_j= \Bigg[ \prod_{m=1}^{j-1} \hat \sigma_m^{z} \Bigg] \hat \sigma_j^{-}~,
\end{equation}
where $\hat{\sigma}_j^{\pm}\equiv( \hat{\sigma}_j^{x} \pm i \hat{\sigma}_j^{y})/2$.

Applying such transformation to the model in Eq.~\eqref{eq:HamBSupp}, 
one has to distinguish three cases~\cite{alvarez17S}:\\
$\bullet$ All indices are different ($i\neq j\neq k\neq l$). In this case
\begin{equation}
  \hat{c}^\dagger_i \, \hat{c}^\dagger_j \, \hat{c}_k \, \hat{c}_l =
  \beta \Bigg[ \prod_{\xi=\zeta_1+1}^{\zeta_2-1} \hat \sigma^z_\xi \Bigg] \Bigg[ \prod_{\xi^\prime=\zeta_3+1}^{\zeta_4-1} \hat \sigma^z_{\xi^\prime} \Bigg] \,
  \hat {\sigma}^{+}_i \, \hat {\sigma}^{+}_j \, \hat {\sigma}^{-}_k \, \hat {\sigma}^{-}_l~,
  \label{eq:HamBSPIN}
\end{equation}
where $\{\zeta_1,\zeta_2,\zeta_3,\zeta_4\}=\{i,j,k,l\}$ are the four reordered indices,
such that $\zeta_1<\zeta_2<\zeta_3<\zeta_4$, and $\beta = {\rm sign}(i-j) \, {\rm sign}(k-l)$;\\
$\bullet$ Two indices are equal (e.g.~$j=l$ and $i\neq j\neq k$). In this case:
\begin{equation}
  \hat{c}^\dagger_i \, \hat{c}^\dagger_j \, \hat{c}_j \, \hat{c}_k =
  \hat \sigma^{+}_{i} \Bigg[ \prod_{\xi=\zeta_1}^{\zeta_2-1} \hat \sigma^{z}_{\xi} \Bigg] \, \hat \sigma^{+}_{j} \, \hat \sigma^{-}_j \, \hat \sigma^{-}_k~,
  \label{eq:HamBPHSJW11}
\end{equation}
where $\{\zeta_1,\zeta_2 \}=\{i,k\}$ are reordered such that $\zeta_1<\zeta_2$;\\
$\bullet$ Indices are equal in pairs (e.g.~$j=k$ and $i=l$). In this case
\begin{equation}
  \hat{c}^\dagger_i \, \hat{c}^\dagger_j \, \hat{c}_j \, \hat{c}_i = \hat \sigma^{+}_{i} \, \hat \sigma^{+}_{j} \, \hat \sigma^{-}_j \, \hat \sigma^{-}_i~.
  \label{eq:HamBPHSJW13}
\end{equation}

As we mentioned in the main text, in order to enforce PHS, one needs to add extra terms of the form $ \hat{c}^\dagger_i \hat{c}_k$ to Eq.~(\ref{eq:HamBSupp}) [cf.~(5) in the main text].
We can again use the JW transformation in order to write each of these one-body contributions in terms of spin-$1/2$ operators:
\begin{eqnarray}
 \hat{c}^\dagger_i \hat{c}_k = \hat \sigma^{+}_{i} \Bigg[ \prod_{\xi=\zeta_1}^{\zeta_2-1} \hat \sigma^{z}_{\xi} \Bigg] \hat \sigma^{-}_k~,
  \label{eq:HamBPHSJW2}
\end{eqnarray}
where $\{\zeta_1,\zeta_2 \}=\{i,k\}$ are reordered such that $\zeta_1<\zeta_2$.

Once the Hamiltonian is written in the spin-$1/2$ representation (spin operators do commute on different sites),
one can safely write its matrix representation in the usual computational basis where
the operator $\hat{\sigma}^z_j$ is diagonal.
Notice that, for the b-SYK Hamiltonian [Eq.~(6)], the JW string is not required.

\begin{figure*}[!t]
  \centering
  \vspace{1.em}
  \vspace{1.em}
  \begin{overpic}[width=0.8\columnwidth]{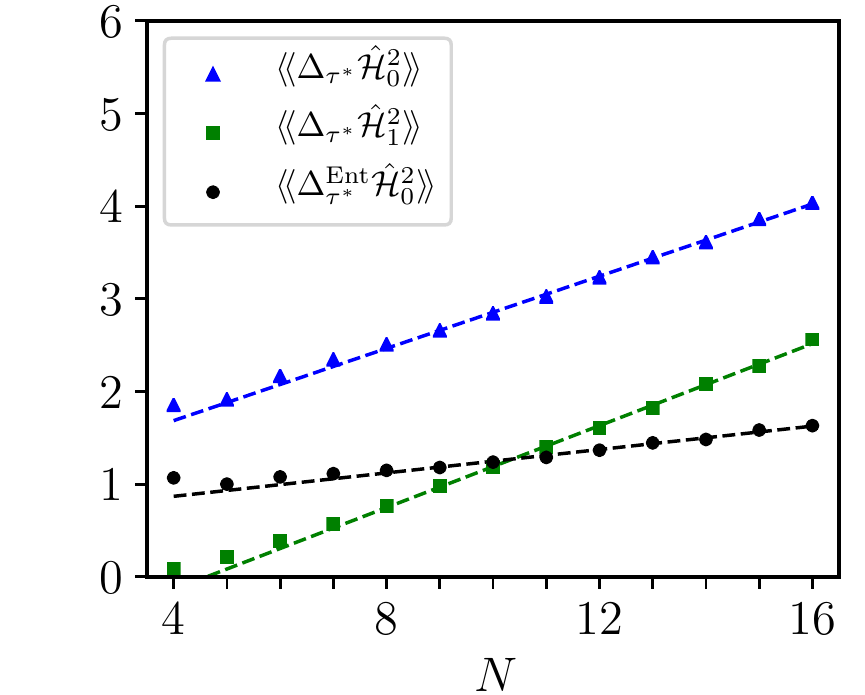}\put(0,80){\normalsize (a)}\end{overpic}
  \begin{overpic}[width=0.8\columnwidth]{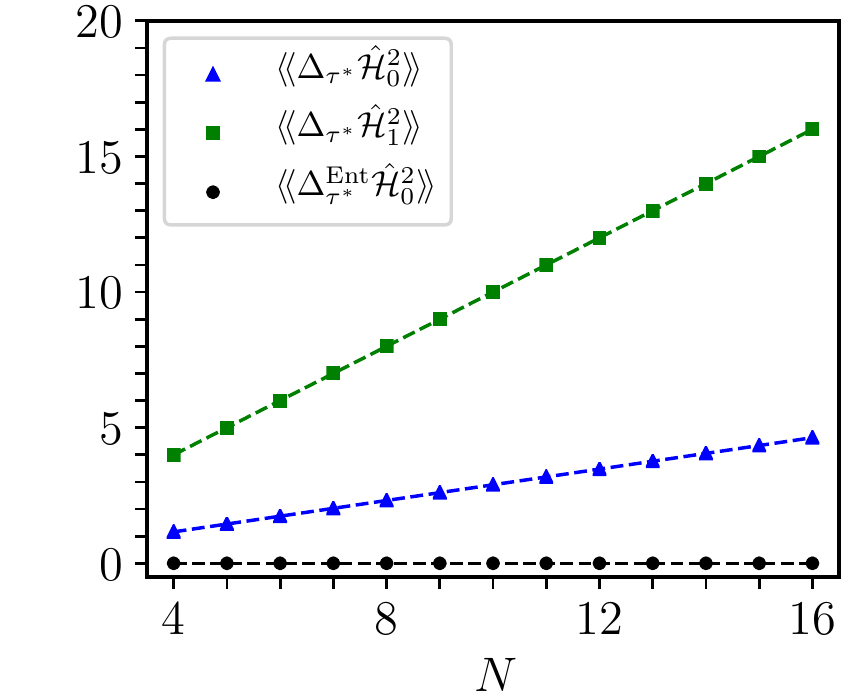}\put(0,80){\normalsize (b)}\end{overpic}\\
  \vspace*{0.2cm}
  \begin{overpic}[width=0.8\columnwidth]{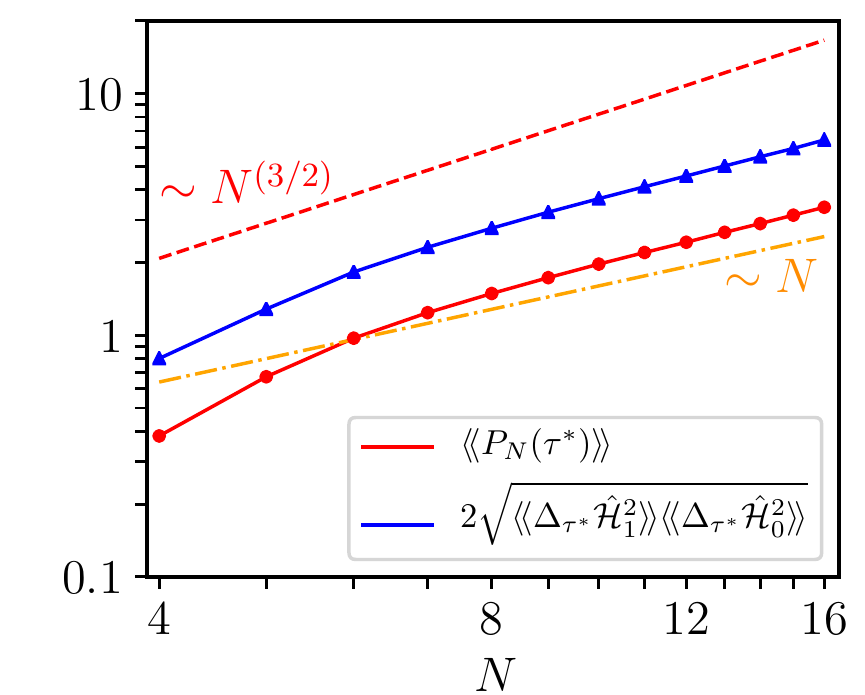}\put(0,80){\normalsize (c)}\end{overpic}
  \begin{overpic}[width=0.8\columnwidth]{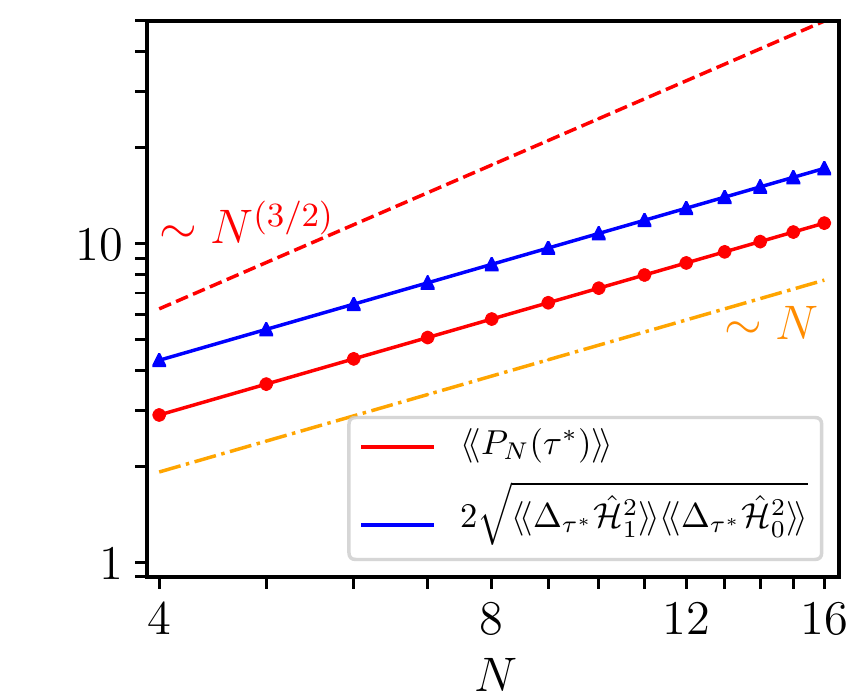}\put(0,80){\normalsize (d)}\end{overpic}\\
  \vspace*{0.2cm}
  \caption{Panels (a,b) The relevant quantities for the bound~(12) in the main text, evaluated at the optimal time $\tau^*$ and averaged over disorder: time-averaged variances $\langle \! \langle \Delta_{\tau^*} \mathcal{\hat H}^2_0 \rangle \! \rangle$ (blue triangles), $\langle \! \langle \Delta_{\tau^*} \mathcal{\hat H}^2_1 \rangle \! \rangle$ (green squares), $\langle \! \langle \Delta^{\rm Ent}_{\tau^*} \mathcal{\hat H}^2_0 \rangle \! \rangle$ (black circles), as functions of $N$. Dashed lines denote linear fits to the numerical results. The four data points corresponding to the smallest $N$ have been always eliminated from the fits. Panels (c,d) The optimal power (red) $\langle\!\langle P_{N}(\tau^*) \rangle\!\rangle$ and the quantity in the right-hand-side of the bound~(12) (blue) are plotted as functions of $N$, in a log-log scale. Dashed lines correspond to power laws $\sim N^{1+k}$ ($k=0.5$: red; $k=0$: orange) and are plotted as guides to the eye. Data in panels (a,c) refer to the b-SYK QB model. Data in panels (b,d), instead, refer to the  parallel QB model.In panels (a,b), $\langle \! \langle \Delta_{\tau^*} \mathcal{\hat H}^2_0 \rangle \! \rangle$ and $\langle \! \langle \Delta^{\rm Ent}_{\tau^*} \mathcal{\hat H}^2_0 \rangle \! \rangle$ are measured in units of $\omega^2_{0}$, while $\langle \! \langle \Delta_{\tau^*} \mathcal{\hat H}^2_1 \rangle \! \rangle$ is measured in units of $J^2$ for both b-SYK and parallel-charging protocols. Data in panels (c,d) are in units of $\omega_{0} J$ for both b-SYK and parallel-charging protocols. This implies that choices need to be made for  the parameters $\bar{J}$ and $K$, in units of $J$: data in this figure have been obtained by setting $\bar{J} = J$ and $K =~J$.  Here and in Fig.~\ref{fig:PowerR}, data for both types of SYK models have been obtained after averaging over $N_{\rm dis} = 10^3$ (for $N=4, \ldots, 10$), $5 \times 10^2$ (for $N=11, 12$), and $10^2$ (for $N=13, \ldots, 16$) instances of disorder in the couplings $\{ J_{i,j,k,l} \}$ and $\{ \bar{J}_{i,j,k,l} \}$.}
  \label{fig:S1:Power}
\end{figure*}

In order to evaluate the properties of the time-evolved state during our charging protocol ($\hbar=1$),
\begin{equation}
  | \psi(\tau) \rangle = e^{- i \mathcal{\hat H}_1 \tau} \bigg( \bigotimes_{j=1}^{N} \ket{\downarrow^{(y)}}_{j} \bigg)~,
\end{equation}
we numerically integrated the equation of motion for $|\psi(\tau)\rangle$ using a fixed-stepsize fourth-order Runge-Kutta method.
To ensure convergence, typical integration time steps of order $\delta t \approx 10^{-3}$ (in units of $1/J$) were used. We checked that our choice of $\delta t$ is always conservative (i.e., it guarantees convergence in time of all our results, within an error bar that is negligible on the scale of the figures).

\section{Derivation of Eq.~(9) in the main text}
From the Heisenberg equation of motion for $0\leq t\leq \tau$ we get:
\begin{equation}\label{eq:S1}
\bigg( \frac{dE_N(t)}{dt} \bigg)^2 = \big| \braket{[\hat{\mathcal{H}}_0,\hat{\mathcal{H}}_1]}_t \big|^2~.
\end{equation}
The Schr\"odinger-Robertson (SR) inequality~\cite{RobertsonS}  yields:
$|\braket{[\hat{\mathcal{H}}_0,\hat{\mathcal{H}}_1]}_t|^2 \leq 4 \, (\delta_t \hat{\mathcal{H}}^2_0 ) \, (\delta_t \hat{\mathcal{H}}^2_1)$,
where $\delta_t \hat{\mathcal{H}}^2\equiv  \braket{ \hat{\mathcal{H}}^2}_t - \braket{ \hat{\mathcal{H}}}^2_t$.
Taking the square root of Eq.~\eqref{eq:S1}, using the SR inequality, applying the integral $\int_0^\tau dt/ \tau$ to both members of Eq.~\eqref{eq:S1},  and using $E_N(0)=0$, we finally get the inequality:
\begin{equation}\label{eq:S2}
P_N(\tau) \equiv \frac{E_N(\tau)}{\tau} \leq 2 \int_0^\tau \frac{dt}{\tau} \sqrt{ (\delta_t \hat{\mathcal{H}}^2_0) \, (\delta_t\hat{\mathcal{H}}^2_1)}~.
\end{equation}
Using the Cauchy-Schwarz inequality with respect to the scalar product induced by $\int_0^\tau dt/ \tau$,
we finally get Eq.~(9) in the main text, i.e.
\begin{equation}
  P_N(\tau) \leq 2 \sqrt{ \Delta_{\tau}\hat{\mathcal{H}}^2_0 \, \Delta_{\tau}\hat{\mathcal{H}}^2_1}~.
  \label{eq:S3}
\end{equation}
  Since the evolution is generated by the charging Hamiltonian $\hat{\mathcal{H}}_1$ itself, the time-average $\int_0^\tau dt/ \tau$ involved in the second term in the r.h.s of Eq.~\eqref{eq:S3} can be computed trivially as
  \begin{equation}
    \Delta_{\tau}\hat{\mathcal{H}}^2_1 = \braket{\mathcal{\hat H}^2_1}-\langle \mathcal{\hat H}_1 \rangle^2.
  \end{equation}
\section{Comparison between quantum and classical many-body batteries}
Consider the bound in Eq.~\eqref{eq:S3} [Eq.~(9) in the main text]. 
 In order to ensure thermodynamic consistency, the average value of the charging Hamiltonian is required to be extensive, $\braket{ \mathcal{\hat H}_1}\sim N$, while its standard deviation, $\big[{\braket{\mathcal{\hat H}^2_1}-\langle \mathcal{\hat H}_1 \rangle^2}\big]^{1/2}$ should scale as $\sqrt{N}$. 
This ensures that relative fluctuations, $\braket{ \mathcal{\hat H}_1}/\big[{\braket{\mathcal{\hat H}^2_1}-\langle \mathcal{\hat H}_1 \rangle^2}\big]^{1/2}$, drop to zero as $N$ goes to infinity, implying the equivalence of all the thermodynamic ensembles (microcanonical, canonical, and grand canonical). This constraint forces $\Delta_{\tau}\hat{\mathcal{H}}^2_1$ to scale at most linearly with $N$.
In the main text, we have ensured thermodynamic consistency of the SYK QB by choosing the appropriate scaling~\cite{Fu16S} with $N$ of the variance $\langle \! \langle J^2_{i,j,k,l} \rangle \! \rangle = J^2/N^3$ of the coupling parameters.

Exploiting the locality of $\mathcal{\hat H}_0 = \sum_j \hat h_j$, with $\hat h_j = \omega_0 \hat \sigma^y_j / 2$,
the first term in the r.h.s.~of Eq.~\eqref{eq:S3} can be written as the sum of two contributions: $\Delta_\tau \mathcal{\hat H}^2_0=\Delta^{\rm Loc}_\tau \mathcal{\hat H}^2_0+\Delta^{\rm Ent}_\tau \mathcal{\hat H}^2_0$, see Eqs.~(10, 11) in the main text, 
\begin{eqnarray}
  \label{eq:BoundEntS}
  \Delta^{\rm Loc}_\tau \mathcal{\hat H}^2_0 & \equiv &
  \frac{1}{\tau} \int_0^\tau dt \sum_i \Big[ \braket{\hat h_i^2}_t - \braket{\hat h_i}^2_t \Big]~, \label{eq:DeltaH0S} \\
  \Delta^{\rm Ent}_\tau \mathcal{\hat H}^2_0~ & \equiv & \frac{1}{\tau} \int_0^\tau dt
  \sum_{i\neq j} \Big[ \braket{\hat h_i \hat h_j}_t - \braket{\hat h_i}_t \braket{\hat h_j}_t \Big], \quad \label{eq:DeltaEntH0S}
\end{eqnarray}
where averages $\langle \, \cdot \, \rangle_t$ are done on the state $|\psi(t)\rangle$ at time $t$.
The first term, being a sum of $N$ factors, is extensive with $N$ by construction. The second term can, in principle, scale quadratically with $N$ if correlations between different sites are established during the dynamics. Here we argue that such correlations have a quantum origin.

Indeed, consider the correlation term
\begin{equation}
  C_{\phi} = \sum_{i\neq j} \Big[ \braket{\hat h_i \hat h_j}_\phi
    - \braket{\hat h_i}_\phi \braket{\hat h_j}_\phi \Big] 
\end{equation}
inside the integral in Eq.~\eqref{eq:DeltaEntH0S}, where averages $\langle \, \cdot \, \rangle_\phi$ are done over a given state $\ket{\phi}$.
Evaluating it on the highly nonclassical Greenberger-Horne-Zeilinger (GHZ) state \cite{NielsenChuangS}, 
\begin{equation}
  \ket{\rm GHZ} = \frac{1}{\sqrt{2}} \bigg( \bigotimes_{j=1}^{N} \ket{\downarrow^{(y)}}_{j}
  + \bigotimes_{j=1}^{N} \ket{\uparrow^{(y)}}_{j} \bigg) \,,
\end{equation}
would result in a quadratic scaling with $N$, i.e., $C_{\rm GHZ} = N(N-1) \, \omega_0^2$.
Conversely such correlation term evaluated over a separable state, $\ket{\phi}=\bigotimes_{j=1}^{N} \ket{\varphi}_{j}$  ($\ket{\varphi}_{j}$ being an arbitrary local state), would trivially vanish. This means that, in order to have a super-linear scaling in the contribution $\Delta^{\rm Ent}_{\tau^*} \mathcal{\hat H}^2_0$, the system has to evolve through highly nonlocal states, as the GHZ state, during the dynamics.  
By definition, classical systems do not build up any entanglement during the charging dynamics, therefore different battery units are uncorrelated $\braket{\hat h_i \hat h_j}_t = \braket{\hat h_i}_t \braket{\hat h_j}_t$ and $\Delta^{\rm Ent}_{\tau^*} \mathcal{\hat H}^2_0 =0$, meaning that the power scales linearly with $N$.

In conclusion, in classical many-body batteries the term $\Delta_{\tau^*} \mathcal{\hat H}^2_0$ scales at most linearly with $N$, while the term $\Delta_{\tau^*} \mathcal{\hat H}^2_1$ is constrained to scale linearly by thermodynamic consistency. 
On the other hand, in quantum many-body batteries the entanglement production enables $\Delta_{\tau}\hat{\mathcal{H}}^2_0$ to scale quadratically with $N$, which in turn implies that the power may scale at most as $N^{3/2}$.

A more detailed analysis of the relation between power and entanglement is given in Ref.~\cite{Farre18S}, which shows that a finite fraction of quantum cells are required to be entangled in a GHZ-like state in order to imply a superextensive charging power.

\section{Power and bounds for the b-SYK and the parallel-charging models}
In the main text it has been shown that a QB charged through the c-SYK model
is able to outperform any classical battery, since 
both $\langle \! \langle \Delta_{\tau^*} \mathcal{\hat H}^2_0 \rangle \! \rangle$
and $\langle \! \langle \Delta^{\rm Ent}_{\tau^*} \mathcal{\hat H}^2_0 \rangle \! \rangle$
grow quadratically with $N$ (see Fig.~3 in the main text).
Time fluctuations of $\mathcal{\hat H}_0$ are thus super-extensive.
On the other hand, as expected,
$\langle \! \langle \Delta_{\tau^*} \mathcal{\hat H}^2_1 \rangle \! \rangle$ is extensive in $N$.
This suggests that the bound~(12), as well as
the optimal power, scale as $N^{3/2}$:
\begin{equation}
\langle\!\langle P_{N}(\tau^*) \rangle\!\rangle \sim N^{1+\frac{1}{2}}~\quad \mbox{(for the c-SYK model)}~,
\end{equation}
a fact that is fully confirmed by our numerical calculations.
In Fig.~\ref{fig:S1:Power}, we show the same quantities for the b-SYK model
[panels (a)-(c)] and for the parallel model [panels (b)-(d)].
It is evident that, in both cases, all of the above mentioned time-averaged variances,
as well as the optimal charging power, grow linearly in $N$,
\begin{equation}
  \langle\!\langle P_{N}(\tau^*) \rangle\!\rangle \sim N \quad \mbox{(for the b-SYK \&  parallel models)}~.
\end{equation}
This rules out the possibility to have a genuine quantum speed-up in the charging process, by using the b-SYK and parallel-charging Hamiltonians.

\section{Comparison between the c-SYK, b-SYK and the parallel-charging models using renormalized Hamiltonians}

{The charging performances of the various quantum batteries (QBs) can be tested by analyzing
the scaling of the optimal power $P_N(\tau^*)$ with the number $N$ of cells [see Eq.~(3) in the main text].
Comparisons between the different models need to be made with great care. We note that the time-evolution operator is $\hat{U}(t) \equiv \exp{(-i \hat{\cal H}_{1} t)}$. The charging Hamiltonian $\hat{\cal H}_{1}$ contains an energy scale, i.e.~$J$ ($\bar{J}$) for the c-SYK (b-SYK) model and $K$ for the parallel-charging model. Here it is important to (i) rule out trivial power enhancements determined by an increase in the energy scale, i.e.~obtained by multiplying the energy couplings by a factor $\alpha>1$, and (ii) compare the three models in a fair manner---``fair" in the sense that, trivially, a parallel charging protocol with $K\geq J, \bar{J}$, for example, may outperform c-SYK and b-SYK charging protocols, and we want to avoid that. 

To rule out these spurious effects, we consider the rescaled charging Hamiltonians~\cite{Binder15S,Campaioli17S}, 
\begin{equation}\label{eq:HamResc}
\hat{\mathscr{H}}_1 \equiv \frac{\mathcal{\hat H}_{1}}{\Vert \mathcal{\hat H}_{1} \Vert}~,
\end{equation}
where $\Vert \hat{\mathcal O} \Vert = \mu_{\mathcal{\hat O}}$ defines the norm of the operator
$\hat{\mathcal{O}}$, $\mu_{\mathcal{\hat O}}$ being its highest singular value. For the sake of convenience
and without loss of generality, we also set to zero the ground-state energy $\epsilon_0$
of all Hamiltonian operators $\mathcal{\hat H}$, by adding a suitable constant.
The charging Hamiltonian (\ref{eq:HamResc}) allows a fair comparison between different QB models. 
In Fig.~\ref{fig:PowerR}, we report the optimal charging power
$\langle\!\langle P_{N}(\tau^*) \rangle\!\rangle$ as a function of $N$, calculated for the c-SYK, b-SYK, and parallel rescaled charging Hamiltonians. 
For the case of the c-SYK and b-SYK models, data have been obtained after averaging over many disorder realizations. Results in this figure are independent of the microscopic energy scale appearing in $\mathcal{\hat H}_1$. 

We see that the c-SYK is the only model for which $\langle\!\langle P_{N}(\tau^*) \rangle\!\rangle$ clearly increases with $N$, thereby presenting a qualitative advantage over the b-SYK and parallel-charging QBs. Concerning the b-SYK QB, its poor performance with respect to its fermionic cousin, the c-SYK QB, indicates that random pair hopping, which both models share, is not enough to guarantee a quantum advantage. The non-local JW strings for fermions are crucial, as they maximize entanglement production during the time evolution and therefore correlations between the $N$ quantum cells.
\begin{figure}[t]
\vspace{1.em}
\begin{overpic}[width=0.9 \columnwidth]{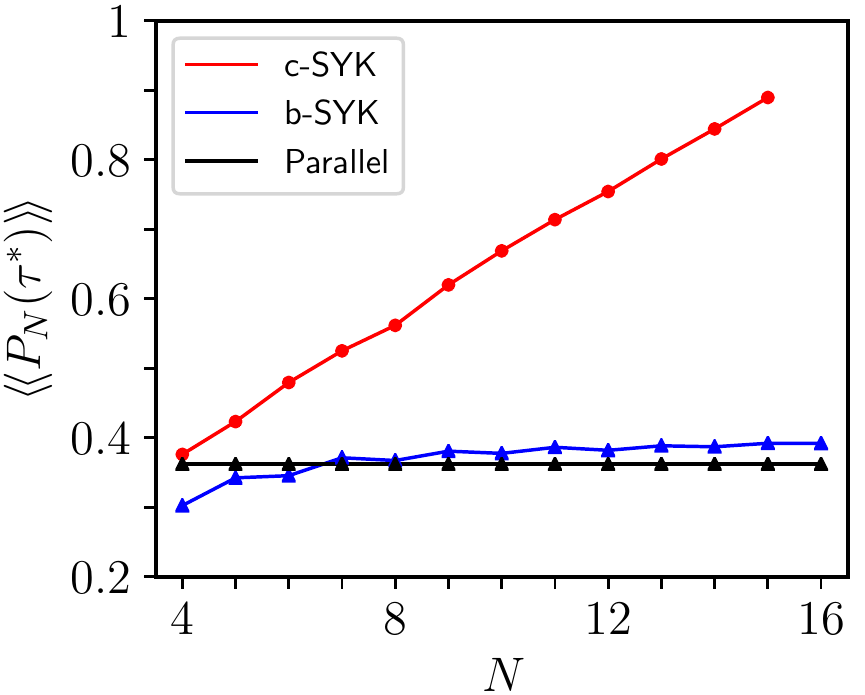}\put(6,73){ }\end{overpic}\caption{
  The dependence of the averaged optimal charging power $\langle\!\langle P_{N}(\tau^*) \rangle\!\rangle$ on the number $N$ of quantum cells, using the rescaled Hamiltonian~\eqref{eq:HamResc}. The averaged optimal charging power shown in this plot is thus measured in units of $\omega_{0}$. In red, we show the optimal power calculated for the c-SYK model with PHS. In blue (black) we show the same quantity for the b-SYK (parallel) model.}
\label{fig:PowerR}
\end{figure}
Note that there is no contradiction between the scaling of the optimal charging power shown in Fig.~\ref{fig:PowerR} for the c-SYK charging protocol and the $\sim N^{3/2}$ scaling seen in Fig.~3(b) in the main text. The point is that, in the former, the rescaled Hamiltonian (\ref{eq:HamResc}) was used. We have checked that the ratio between the two optimal charging powers yields the correct bandwidth of the c-SYK model, which scales linearly with $N$.}

\section{Asymptotic dynamics of the SYK model and random states}

The SYK model is known to exhibit peculiar properties such as non-integrability, the absence of any local conserved
quantity, and quantum chaos~\cite{Liu18S,Maldacena16S}. Such properties imply that any time-evolved state at long times can be locally approximated  by a suitable thermal state. We therefore expect that in the c-SYK QB charging protocol,
after an initial transient time, the population $p_k (\tau)$ of the energy levels of the Hamiltonian
$\mathcal{\hat H}_0$ becomes independent of $\tau$. More precisely, we expect it to be well approximated
by that of a random state in the $2^N$-dimensional Hilbert space.
In fact, such time independence of $p_k$ for the c-SYK model is clearly visible in Fig.~2(a) of the main text, already for $\tau \gtrsim 1$ (in units of $1/J$).
Panel (b) of the same figure suggests that a similar situation may also occur for the b-SYK model as well, but at comparatively longer times.
\begin{figure}[t]
\begin{overpic}[width=0.8\columnwidth]{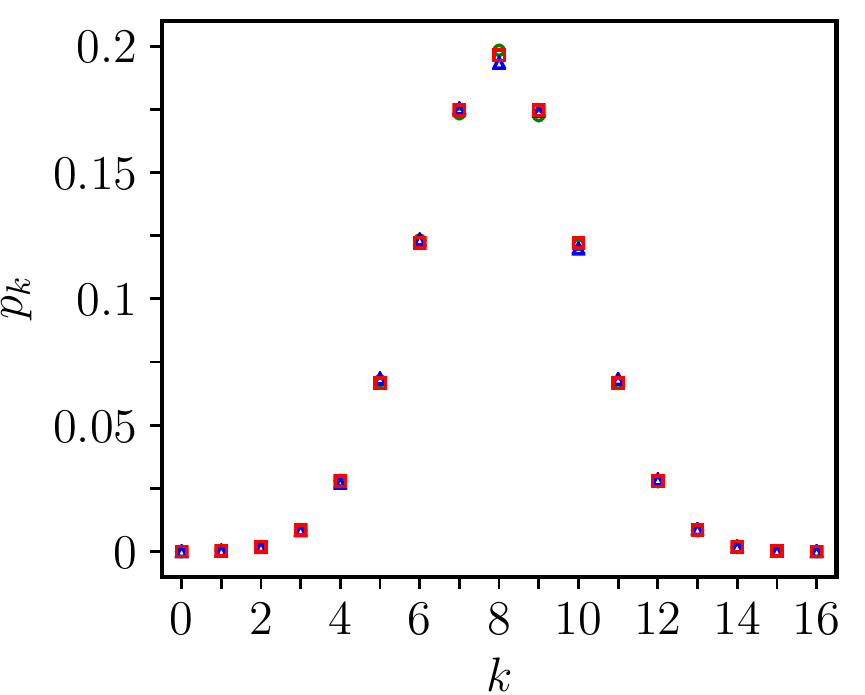}\put(0,80){\normalsize }\end{overpic}\caption{The energy-level population (blue triangles) $p_k$ as a function of $k$, for the c-SYK model, evaluated numerically at $\tau=4$ (in units of $1/J$). These numerical results are compared against the analytical result in Eq.~(\ref{eq:binomial}) (red squares) and the prediction based on a random state (green circles). The two latter outcomes (red and green symbols) turn out to be indistinguishable from the numerically obtained data (blue symbols).\label{fig:S2:Pk}}
\end{figure}

This is quantitatively analyzed in Fig.~\ref{fig:S2:Pk}, where we report the same data
of Fig.~2(a), once the time is fixed to $\tau = 4$, in units of $1/J$ (blue triangles).
One can immediately recognize that such distribution of energy levels agrees nearly perfectly with the one
corresponding to a completely mixed state, $\rho^{\rm (r)} = \mathbb{1}/2^N$ (red squares),
The latter is simply given by a binomial distribution
\begin{equation}\label{eq:binomial}
  p^{\rm (r)}_k = \frac{1}{2^N} \binom{N}{k}~.
\end{equation}
A very similar result can be obtained if a random pure state, $|\psi^{\rm (r)}\rangle = \sum_n c_n |n\rangle$, is taken, $c_n$ being complex numbers with randomly distributed amplitude and variance and satisfying $\sum_n |c_n|^2 = 1$ (green circles).

This reasoning hints at a fast thermalization of the c-SYK model to an infinite temperature state.

\section{Charging power of a Dicke battery}
Unlike for the c-SYK battery, the power of a Dicke battery~\cite{Ferraro17S} does not exhibit a super-linear scaling, provided consistency with the thermodynamic limit is correctly enforced~\cite{Farre18S}.
Dicke batteries are unitarily charged via a protocol that is slightly different from the one described in Eq.~(2) of the main text. In fact, both the charging system A and the battery B are quantum mechanically described by the time-dependent Hamiltonian
\begin{equation}\label{eq:protocolS}
  \hat{\mathcal{H}}(t)=\hat{\mathcal{H}}_{\rm A}+ \hat{\mathcal{H}}_{\rm B}+\lambda(t)\hat{\mathcal{H}}_{\rm int}~.
\end{equation}
where $\hat{\mathcal{H}}_{\rm A}$ ($\hat{\mathcal{H}}_{\rm B}$) is the free Hamiltonian acting on the system A (B),  $\lambda(t)$ is a classical control parameter, which is assumed to be equal to one if $t\in [0,\tau]$ and zero elsewhere, and $\hat{\mathcal{H}}_{\rm int}$ is an interaction Hamiltonian which couples the charging system and the battery, thus enabling the charging process to occur.
A Dicke battery is made by $N$ qubits (the battery cells) charged by a cavity mode.
The Hamiltonian terms in Eq.~\eqref{eq:protocolS} are given by
\begin{eqnarray}
  \label{eq:DickeHS}
  \hat{\mathcal{H}}_{\rm A}&=& \omega_0 \hat{a}^\dagger\hat{a}~,\\
  \hat{\mathcal{H}}_{\rm B}&=&\frac{\omega_0}{2} \bigg(\sum_{i=1}^N \hat{\sigma}^z_i+ \frac{N}{2}\bigg)~,\\
  \label{eq:DickeHSInteraction}
  \hat{\mathcal{H}}_{\rm int}&=&\frac{g}{\sqrt{N}} \big( \hat{a}^\dagger+\hat{a}\big) \sum_{i=1}^N \hat{\sigma}^x_i~,
\end{eqnarray}
where $\hat{a}$ ($\hat{a}^\dagger$) is a bosonic annihilation (creation) operator, $\omega_0$ is the characteristic frequency of both subsystems, and $g$ the coupling strength. The prefactor $1/\sqrt{N}$ ensures the thermodynamic consistency of the model \cite{Farre18S}.
As a matter of fact, the interaction Hamiltonian $\hat{\mathcal{H}}_{\rm int}$ can be derived from a dipole light-matter interaction of the form $\sum_{i=1}^N \hat{\bm{d}}_i\cdot \hat{\bm{E}}_i$, where $\hat{\bm{d}}_i$ is the dipole operator acting on the $i$-th cell and $\hat{\bm{E}}_i$ is the cavity electric field evaluated at the position $i$. While each dipole operator $\hat{\bm{d}}_i$ does not carry any scaling with $N$, the electric field inside a cavity with volume $V$ scales like $1/\sqrt{V}$. In a cavity system, the correct thermodynamic limit consists in considering $N\to \infty$, $V\to \infty$, with $N/V\to {\rm const}$. 
This means that $V \sim N$. Thus $1/\sqrt{N}$ is the correct prefactor to enforce a well-defined thermodynamic limit.

\begin{figure}[b]
\begin{overpic}[width=0.8\columnwidth]{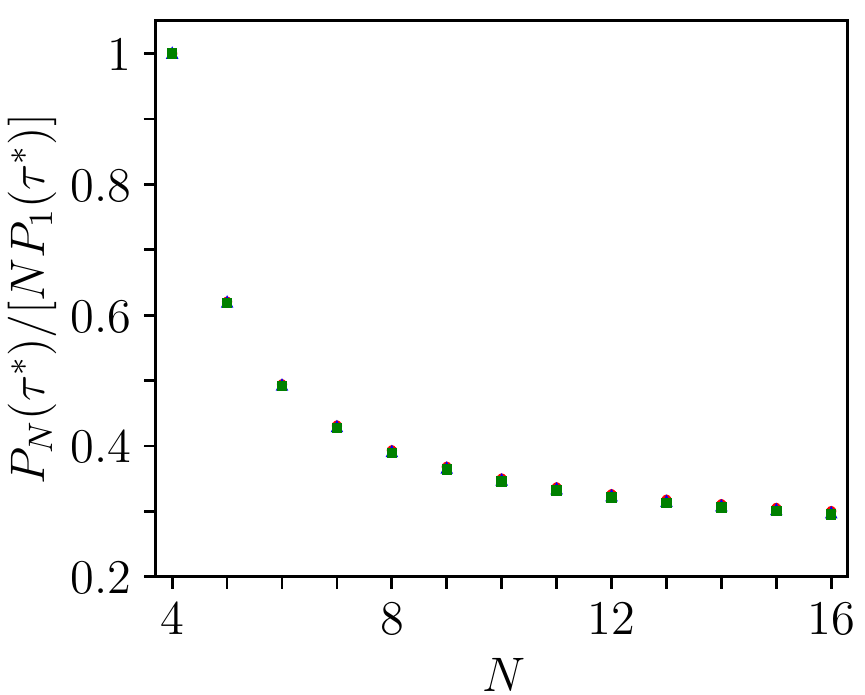}\put(0,80){\normalsize}\end{overpic}\caption{The optimal charging power $P_N(\tau^*)$, in units of $N P_1(\tau^*)$, as a function of $N$. Different symbols and colors refer to various values of the coupling strengh: $g/\omega_0=0.05$ (red circles), $g/\omega_0=0.5$ (blue triangles), $g/\omega_0=2$ (green squares).\label{fig:SDickePower}}
\end{figure}

At time $t=0$, the charger is prepared in a $N$-photon Fock state, while the qubits are prepared in their ground
state, namely
\begin{equation}\label{eq:initialState}
  \ket{\psi(0)}=\ket{N}_{\rm A} \otimes \ket{0}_{\rm B}~.
\end{equation}
The energy injected in the $N$ cells of the Dicke battery and the corresponding average power are defined as
\begin{eqnarray}
\label{eq:energypowerS}
E_N(\tau)&\equiv & {\rm tr}\big[ \hat{\mathcal{H}}_{\rm B} \, \rho_{\rm B}(\tau)\big]~,\\
P_N(\tau)&\equiv &\frac{E_N(\tau)}{\tau} ~. 
\end{eqnarray}
The optimal charging time $\tau^*$ is defined as in Eq.~(3) in the main text, $P_N({\tau}^*)=\max_{\tau>0} P_N({\tau})$.
In Fig.~\ref{fig:SDickePower}, we explicitly show the optimal charging power $P_N(\tau^*)$, normalized by $N$ times the optimal charging power of a single cell $P_1(\tau^*)$. This ratio tends to saturate to a constant for $N$ large enough, meaning that the power of a Dicke battery does not display a super-linear scaling.


\begin{thebibliography}{77}
\bibitem{majorana}
  We have decided to work with complex rather than Majorana fermions~\cite{Kitaev15}, since experimental realizations of the latter require complicated setups, including e.g.~spin-orbit coupling, magnetic fields, and proximity superconductivity.
\bibitem{rescaling}
  This choice is not restrictive. Indeed, results for other values of parameters can be obtained through the following rescaling procedure, without the need of carrying out further simulations. A linear rescaling of the QB energy scale, $\omega_0\to \alpha \omega_0$, reflects into a linear change of the power, i.e.~$P_N(\tau)\to \alpha P_N(\tau)$. The same occurs under a linear rescaling of the charging Hamiltonian coupling constant, $J \to \alpha J$.
\end{thebibliography}
\end{document}